\documentclass[a4paper,11pt]{article}
\usepackage{pos}

\title{Study of blazars detected by Fermi-LAT as high-energy neutrino sources.}
 \ShortTitle{Fermi-LAT Blazars as neutrino progenitors.}

\author*[a]{Antonio Galv\'an}
\author[a]{Nissim Fraija}
\author[a]{Edilberto Aguilar Ruiz}
\author[a]{Jose Antonio de Diego Onsurbe}
\author[b,c,d]{Maria G. Dainotti}
\affiliation[a]{Instituto de Astronom\' ia, Universidad Nacional Aut\'onoma de M\'exico, Circuito Exterior, C.U., A. Postal 70-264, 04510 M\'exico City, M\'exico}

\affiliation[b]{National Astronomical Observatory of Japan, 2-21-1 Osawa, Mitaka, Tokyo 181-8588, Japan}
\affiliation[c]{Space Science Institute, Boulder, CO, USA}
\affiliation[d]{The Graduate University for Advanced Studies, SOKENDAI, Shonankokusaimura, Hayama, Miura District, Kanagawa 240-0193, Japan}

\emailAdd{agalvan@astro.unam.mx}

\abstract{Besides the neutrino source detected by IceCube, NGC 1068, the association of the IceCube-170922A neutrino with the blazar in a flaring state among several wavelengths (from radio up to high-energy (HE) gamma-rays), the site and mechanisms of production of HE neutrino remains in discussion. Extragalactic sources such as Quasars, Blazars, Radio galaxies, and Gamma-ray bursts have been proposed as progenitors of HE neutrinos. In this work, we study the Blazars reported by Fermi-LAT in the 4LAC catalog, which are embedded inside the 90\% error of the best-fit position from the neutrinos reported by IceCube. We propose a one-zone lepto-hadronic scenario to describe the broadband Spectral Energy Distribution and then estimate the number of neutrinos to compare with those in the direction of each source. A brief discussion is provided of the results.}

\ConferenceLogo{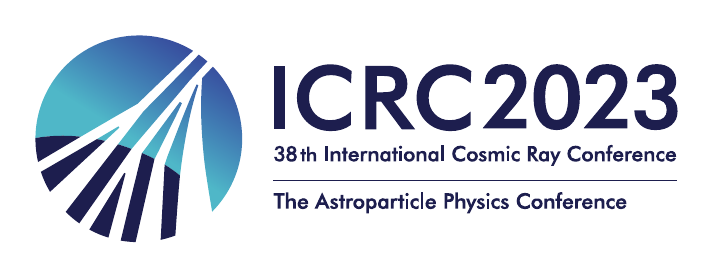}

\FullConference{%
38th International Cosmic Ray Conference (ICRC2023)\\
  26 July - 3 August, 2023\\
  Nagoya, Japan}

\begin{document}
\maketitle

\section{Introduction}

Over a century has passed since Victor Hess's first discovery of Ultra High Energy Cosmic Rays (UHECRs), yet their origin remains unknown. Because of how CRs and magnetic fields interact, pinpointing where precisely a CR was born is difficult \cite{COUTU20121355}. The intensity of $\gamma$-rays may be attenuated by interactions with Cosmic Microwave Background (CMB) photons \citep[e.g.][]{2012MNRAS.422.3189G, 2017A&A...603A..34F} by a quantity that relies on both the $\gamma$-rays' redshift and their energy. Neutrinos travel unfettered across magnetic fields from their source to Earth because they have no net electric charge \cite{2019ARNPS..69..477M}. Neutrinos are unaffected by their weak force or gravitational interactions with matter as they journey through the cosmos. High Energy neutrinos may obscure the hunt for extraterrestrial UHCR accelerators.\\

\noindent A variety of theoretical models attempt to account for the source's spectral energy distribution (SED) in Blazars \cite[e.g.][]{2018Sci...361.1378I, 2019ApJ...874L..29R, 2017ApJS..232....7F}. Leptonic models successfully describe the low- and high-energy peaks of Blazar objects without presenting problems for the standard shock acceleration mechanism with one-zone geometry.  Even though theoretical techniques that use leptonic models to understand the SED of Blazars have been effective, it has been shown that the existence of protons in the jet is necessary to interpret the multi-wavelength observations.

\section{Fermi-LAT}
In June of 2008, the Fermi spacecraft was sent into orbit, and ever since then, it has been continuously surveying the sky. The satellite comes with two different types of instruments; Large Area Telescope (LAT) and Gamma-ray Burst Monitor (GBM) \cite{2009ApJ...697.1071A}. The LAT team has published the 4FGL catalog \cite{2020ApJS..247...33A}, which details the sky sources that have been detected emitting photons between 50 MeV and 1 TeV during the telescope's first decade of operation. We consider the 4LAC-DR2 \cite{2020arXiv201008406L} in this study, encompassing the ten years of Fermi-LAT AGN detection data. This survey has 285 gamma-ray emitting active galactic nuclei (AGN), including 39 FSRQ, 59 BL Lacs, 185 Blazar Candidates of unknown type (BCU), and two radio galaxies. The 4LAC is divided into a High Latitude sample consisting of AGNs located at or above a galactic latitude of $|b| \mathrm{>} 10^{\circ}$ and a Low Latitude sample consisting of sources located at or below a galactic latitude of $|b| \mathrm{<} 10^{\circ}$. Of the AGNs observed by LAT, 262 are found in the High Latitude sample, while the Low Latitude sample consists of just 23 sources from the catalog, all BCUs\footnote{The \texttt{FITS} tables for this catalogs are public and are available at \url{https://fermi.gsfc.nasa.gov/ssc/data/access/lat/4LACDR2/}}. In this work, we take into consideration blazars from the High Latitude sample, because they are likely to be connected with neutrinos with very high kinetic energy.

\section{Neutrino Data}

According to the High-Energy Starting Events (HESE) database, the IceCube cooperation has detected 82 HESE-neutrino events reported in \cite{2013Sci...342E...1I, 2014PhRvL.113j1101A, 2015ICRC...34.1081K, 2017ICRC...35..981K}.  Meanwhile, the Extremely High Energy (EHE) catalog recorded 36 neutrinos reported in \cite{2016ApJ...833....3A, 2017ICRC...35.1005H}.  In addition, a real-time warning system was created to provide a fast reaction to electromagnetic transients that may be connected with neutrinos. The system uses the AMON consortium to broadcast HESE and EHE events in real time reported in \cite{2019ICRC...36.1021B}. This work takes into consideration 139 track neutrinos in total.

\section{Theoretical model}\label{ch:Model}

\subsection{One-zone SSC model}
Leptonic scenario one-zone SSC model is used to characterize the broadband SED. In this theory, the magnetic field surrounds the region where an accelerated population of electrons causes an emission.  When photons from synchrotron radiation interact with the parent electron population, they are up-scattered to higher energies through inverse Compton scattering \cite[e.g.][]{2017APh....89...14F, 2020arXiv201101847A}. We adopted the theoretical framework derived by  \cite{2008ApJ...686..181F}. This scenario assumes an electron distribution described by:



\begin{equation}
N_{e}(\gamma^{\prime}) = K_{e}\left[ \left( \frac{\gamma^{\prime}}{\gamma_{break}^{\prime}} \right)^{-p}H(\gamma_{break}^{\prime} - \gamma^{\prime})  + \left( \frac{\gamma^{\prime}}{\gamma_{break}^{\prime}} \right)^{-(p+1)} H(\gamma^{\prime} - \gamma_{break}^{\prime})    \right] H(\gamma^{\prime}; \gamma_{1}^{\prime}, \gamma_{2}^{\prime})\,,
\label{eq:Ne}
\end{equation}

\noindent where, $H$ is the Heaviside function\footnote{$H(x; x_{1}, x_{2}) := 1$ if $x_{1} \le x \le x_{2}$ ; 0 in other case.}. Once selected the electron distribution, the synchrotron spectrum is described by:

\begin{equation}
    f_{syn}(\epsilon) =  \frac{\sqrt{3}\delta_{D}^{4}\epsilon^{\prime}e^{3}B}{4\pi h d_{L}^{2}} \int_{1}^{\infty} N_{e}^{\prime}(\gamma^{\prime})R(x) d\gamma^{\prime},
\label{eq:Synch}  
\end{equation}

\noindent where $\epsilon = h\nu/(m_{e}c^{2})$ is the frequency normalized at the electron mass, $\delta_{D}$ is the Doppler factor, $e$ the fundamental charge, $h$ the Planck's constant, $d_{L}$ the luminosity distance from the source and $R(x)$ is defined in \cite{2008ApJ...686..181F}. In the high-energy regime, the SSC spectrum is described by:

\begin{equation}
  f_{ssc}(\epsilon_{s}) =  \frac{9}{16}\frac{(1+z)^{2}\sigma_{T}\epsilon_{s}^{\prime 2}}{\pi \delta_{D}^{2} c^{2} t_{v, min}^{2}} \int_{0}^{\infty} \frac{f_{syn}(\epsilon)}{\epsilon^{\prime}} d\epsilon^{\prime} \int_{\gamma_{min}^{\prime}}^{\gamma_{max}^{\prime}} \frac{N_{e}^{\prime}(\gamma^{\prime}) }{\gamma^{\prime 2}} F_{c}(q, \Gamma_{e}) d\gamma^{\prime}\,. 
  \label{eq:SSC}
\end{equation}

\noindent where $z$ is the redshift, $\sigma_{T}$ is the Thompson cross section, $t_{var, min}$ is the variability timescale and $F_{C}(q, \Gamma_{e})$ is the Compton cross section \citep{1968PhRv..167.1159J}.

\subsection{Hadronic model}

Accelerated relativistic protons in the jet enhance the possibility that photo-hadronic interactions are responsible for the GeV-TeV $\gamma$-ray emission \cite{romero2013introduction}. Targets for photo-hadronic interactions are the photon field produced when relativistic electrons introduced into the highly magnetized blob lose energy mostly via synchrotron radiation. Inverse Compton scattering of synchrotron photons and $\pi^0$ and  $\pi^\pm$ decay products from hadronic interactions are modelled to account for the high-energy peak in the SED, whereas synchrotron radiation accounts for the low-energy hump. Charged ($\pi^+$) and neutral ($\pi^0$) pions are generated via the following channels \cite{2008PhRvD..78c4013K}:  
 \begin{eqnarray}
p\, \gamma &\longrightarrow&
\left\{
\begin{array}{lll}
p\,\pi^{0}\   &&   \mbox{fraction }2/3, \\
n\,  \pi^{+}      &&   \mbox{fraction }1/3\,.\nonumber
\end{array}\right. \\
\end{eqnarray}
Then,  neutral pion decays in two photons, $\pi^0\rightarrow \gamma\gamma$,  carrying each one $10\%$ of the proton's energy $E_p$.   The efficiency of this processes can be written as
{\small
\begin{equation}\label{eficiency}
f_{\pi^0} =\frac{R_b}{2\gamma^2_p}\int\,d\epsilon\,\sigma_\pi(\epsilon)\,\xi_{\pi^0}\,\epsilon\int dx\, x^{-2}\, \frac{dn_\gamma}{d\epsilon_\gamma} (\epsilon_\gamma=x)\,,
\end{equation}
}
with $dn_\gamma/d\epsilon_\gamma$ the seed-photon spectrum , $\xi_{\pi^0}=0.2$, $\sigma_\pi(\epsilon_\gamma)$ the pion cross section and $\gamma_p$ is the Lorentz factor of protons. Solving Eq. \ref{eficiency}, we have  
{\small
 \begin{eqnarray}
f_{\pi^0} \simeq \frac{\sigma_{\rm p\gamma}\,\Delta\epsilon_{\rm res}\,\xi_{\pi^0}\, L_{\rm \gamma}}{4\pi\,\Gamma^2\,R_b\,\epsilon_{\rm pk,\gamma}\,\epsilon_{\rm res}}
\begin{cases}
\left(\frac{\epsilon^{\pi^0}_{\gamma}}{\epsilon^{\pi^0}_{\gamma,c}}\right)^{\beta_h-1}       &  \epsilon_{\gamma} < \epsilon^{\pi^0}_{\gamma,c}\\
\left(\frac{\epsilon^{\pi^0}_{\gamma}}{\epsilon^{\pi^0}_{\gamma,c}}\right)^{\beta_l-1}                                                                                                                                                                                                            &   \epsilon^{\pi^0}_{\gamma,c} < \epsilon_{\gamma}\,,\\
\end{cases}
 \end{eqnarray}
}
where $\beta_h\approx 2.2$ and $\beta_l\approx 1.2$ are the high and low photon indexes, respectively,  $L_{\rm \gamma}$ is the seed-photon luminosity, $\Delta\epsilon_{\rm res}$=0.2 GeV,  $\epsilon_{\rm res}\simeq$ 0.3 GeV, $\epsilon_{\rm \gamma, pk}$ is the seed-photon energy and  $\epsilon^{\pi^0}_{\gamma,c}$ is the break photon-pion energy, which can be written as 
\begin{equation}
\epsilon^{\pi^0}_{\gamma,c}\simeq 31.87\,{\rm GeV}\, \Gamma^2\, \left(\frac{\epsilon_{\rm \gamma, pk}}{ {\rm MeV}}\right)^{-1}\,.
\label{pgamma}
\end{equation}

\section{Analysis and Results}

IceCube neutrinos and FSRQ detected by Fermi-LAT were physically correlated to find possible matches. We consider the neutrino's location and check to see if the angular gap between a Fermi-LAT source and the neutrino position error establishes a physical relationship. Table \ref{Table:Correlations} lists the eight probable correlations we identify between IceCube neutrinos and FSRQ detected by LAT. We analyze solely the spatial coincidence and calculate the probability that each neutrino is related to each FSRQ. Given that Fermi-LAT's high-latitude 4LAC sample extends to $|b| > 10^{\circ}$, this implies that the 4LAC spans an area of $\mathcal{A} \sim 34088.6 \ \rm{deg}^2$.\\

\noindent We examine neutrinos and the reported error of $\mathcal{E}$ from the IceCube experiment. If a neutrino is connected with a FSRQ inside their error area, then the chance of this happening is $\mathcal{P} = \frac{\mathcal{N}_{\nu} \mathcal{E}}{\mathcal{A}}$. The independent probability for each occurrence to be related is in the final column of table \ref{Table:Correlations}.\\

\begin{table*}
\caption{FSRQ correlated to IceCube observations of neutrino tracks. The 4FGL source name is listed first, followed by the redshift. In the third column, we see the neutrino name, followed by its counterpart, the FSRQ name, in the lower energy bands. The two last columns show the distance from the best-fit position to the gamma-ray source and the probability of an association.}\label{Table:Correlations}
\begin{tabular}{cccccc}
\hline
4FGL Name  & z & Associated & IceCube neutrino & Ang. Sep & $\mathcal{P}$ \\
 &    &  &  & ($\mathrm{deg}$)  & \\
 \hline \hline
J2226.8+0051 &  2.26 & PKS B2224+006 & HESE 44 & 0.68 &  0.018 \\
J1557.9-0001 &  1.77 & PKS 1555+001 & HESE 76 & 0.81 &  0.018\\
J1457.4-3539 &  1.42 & PKS 1454-354 & IceCube-181014 & 1.09 & 0.043\\
J1504.4+1029 &  1.84 & PKS 1502+106 & IceCube-190730A & 0.31 & 0.022 \\
J1858.7+5708 & 0.077 & 87GB 185759.9+570427 & IceCube-191215A & 1.88 & 0.052\\

J1103.0+1157 &  0.91 & TXS 1100+122 & IceCube-200109A & 1.26 & 0.092\\ 
J0206.4-1151 &  1.66 & PMN J0206-1150 & IceCube-201130A & 1.07 & 0.022\\
J2108.5+1434 &  2.02 & OX 110 & IceCube-211216A & 1.61 & 0.052 \\
\hline
\end{tabular}
\end{table*}

\noindent Python was used to implement the model discussed in section \ref{ch:Model}. The chosen fitting method is the Monte Carlo Markov Chain (MCMC) technique implemented in the Python module \texttt{emcee} \citep{2013PASP..125..306F}. Once the initial values are determined, a 512-walker, 12,000-step MCMC sampler is constructed. Magnetic field ($B$), Doppler factor ($\delta_{D}$), minimum, break, and maximum Lorentz factors from the electron population ($\rm{\gamma^{\prime}_{m}}$, $\rm{\gamma^{\prime}_{M}}$), and the shape of a broken power law population with slopes $p$ and normalization constant $\rm{K_e}$ were all inputs to the MCMC procedure. The hadronic component have seed temperatures ($\rm{T}$) were fixed at 2.7, 58.73, and 18509.20 $\rm{K}$, respectively, because the power law index of the proton distribution was fixed at 2, and the target photons were the CMB, infrared, and optical, therefore freeing the constant normalization $\rm{K_{p}}$. The proton characteristic energy $E^{*}$ , when a proton collides head-on with a thermal photon of energy $\rm{kT}$, is selected as the break energy on the proton population. Due to the lack of multiwavelength simultaneous observations in some Quasars, at the neutrino arrival time as in the case of the PKS B2224+006 FSRQ, the SED modeling is taken into consideration the historical emission of each source, with the purpose of constraining the model, mainly in the synchrotron emission.\\

\section{Summary}

We found 27 spatial coincidences between point-like sources reported in 4LAC-DR2 and neutrinos detected by IceCube after considering 13 years of data and performing a cross-match with the AGNs detected by Fermi-LAT over ten years and reported in their 4FGL \cite{2020ApJS..247...33A, 2020arXiv200511208B}. If the coincidence is not exact, the $\gamma$-ray sources are still inside the error position of the neutrino, with a separation angle of less than 2$^{\circ}$ from the neutrino's best-fit position. The potential counterpart described in 4LAC and other catalogs was considered for these AGNs. Only eight sources are left when this study is limited to those with a FSRQ spectrum. The neutrino flare observed in TXS 0506+056 before the detection of IceCube-170922A \cite{2018Sci...361..147I} strongly motivated the consideration of three-time windows for these eight sources: weekly ($\pm$ one week after trigger time), monthly ($\pm$ one month after trigger time), and yearly ($\pm$ one year after trigger time). For each one, the SED was modeled with a lepto-hadronic scenario and computed the theoretical flux from p$\gamma$ interactions taking into consideration seed photons from CMB, infrared and optical population. 

\bibliographystyle{JHEP}
\bibliography{ICRC-2023_Galvan} 

\end{document}